\documentclass[12pt]{article}

\input axodraw.sty
\usepackage{graphics}
\newcommand{\comment}[1]{}
\newcommand{\bea}{\begin{eqnarray}}
\newcommand{\eea}{\end{eqnarray}}
\newcommand{\be}{\begin{equation}}
\newcommand{\ee}{\end{equation}}

\begin{document}

\begin{center}
{\Large \bf \boldmath T-odd observables from anomalous 
$tbW$ couplings in single-top W associated production 
at the LHC}

\bigskip
\bigskip
\bigskip

{\large Saurabh D. Rindani\footnote{email: sdrindani@gmail.com}
}

\bigskip

{\it Theoretical Physics Division, Physical Research Laboratory,\\ 
\vskip .1cm
Navrangpura, Ahmedabad 380009, India}

\end{center}

\bigskip
\bigskip
\centerline{\large Abstract}

\bigskip
We investigate the possibility that an imaginary 
anomalous $tbW$
coupling can be measured in the process $pp \to tW^-X$
by means of T-odd observables.
One such observable is the polarization of the top quark
transverse
to the production plane. Another is the asymmetry between the cross
sections  
with a positive and with a negative value of a certain
T-odd correlation constructed out of observable
momenta when the top quark decays leptonically. The mean value of
this correlation, which is also T-odd, is the third observable.  
These three T-odd
observables are shown to be proportional to the imaginary part of only 
one of the $tbW$ anomalous couplings, the other couplings giving
vanishing contributions. 
This imaginary part could
signal either a CP-odd coupling, or an absorptive part in the effective
coupling. 
We estimate the 1-$\sigma$ limits that might be
derived in the case of each of these observables for centre-of-mass
energies 7, 8, 13 and 14 TeV at the LHC.

\section{Introduction}

The Large Hadron Collider (LHC), a proton-proton collider, 
 has seen several successes in establishing the validity of many aspects of 
the
standard model (SM), and constraining physics beyond the standard model.
 With an
enhanced luminosity, as in the future runs such as the High-Luminosity
version (HL-LHC), results with higher precision are anticipated.
Apart from investigating with great accuracy the parameters of the
standard model (SM), future experiments will also strive to discover
new physics -- either with the discovery of new particles, or
checking if the measured couplings of the SM particles violate
predictions from the SM. 

One of the important constituents of the SM is the top quark, the
heaviest quark known. The properties of the top quark have been 
 studied in
great detail at the LHC, particularly through the most abundant process
of top-pair production. However, some of the electroweak properties of
the top quark can be studied through the single-top production
processes \cite{Giammanco:2015bxk, Andrea:2023yap}.
For example, single-top production enables the determination
of the CKM matrix element $V_{tb}$. Of these processes, the $t$-channel 
process has the largest cross
section, followed by the $tW$ associated production process. The
$s$-channel single-top production process is also possible, though it
has the least cross section of the three 
\cite{Giammanco:2015bxk, Andrea:2023yap}.

While the $tW$ production process was measured earlier at the LHC for
centre-of-mass (cm) 
energies of 7 and 8 TeV, the recent measurements are carried out at the
cm energy of 13 TeV. 
The 
CMS collaboration at the LHC has recently 
 measured the $tW$ differential cross
section at 13 TeV with an integrated luminosity of 138 fb$^{-1}$, with a
central value of 79.2 for the total cross section
\cite{CMS:2022ytw}.
The ATLAS collaboration at the LHC has studied the $tW$ production 
process recently at 13 TeV for a luminosity of 140 fb$^{-1}$ and
obtained a central value of 75 pb for the total cross section  
\cite{ATLAS:2024ppp}.

The $tW$ production process occurs at the lowest order through the 
exchange of either a $t$-channel 
virtual $t$ quark 
or an $s$-channel $b$ quark from  an initial state of a 
gluon and a $b$ quark from the protons, producing 
a top quark
through an anomalous $tbW$ vertex, as shown in the diagrams in Fig. 1.
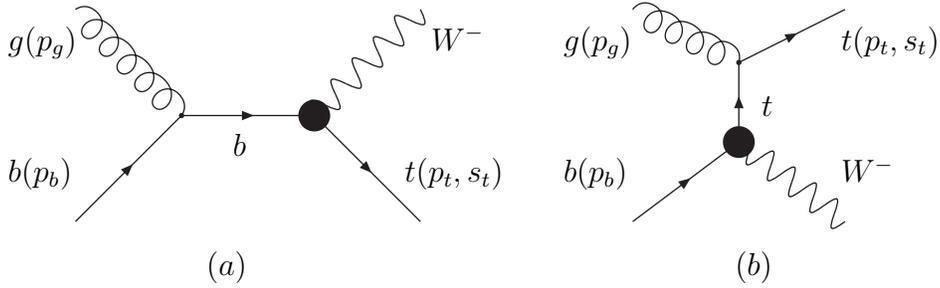
\begin{figure}[hbt]
\begin{center}
\begin{picture}(800,130)(50,0)
\Gluon(100,100)(140,60){5}{5}
\ArrowLine(100,20)(140,60)
\Vertex(140,60){1}
\ArrowLine(140,60)(190,60)
\Vertex(190,60){6}
\Photon(190,60)(230,100){5}{5}
\ArrowLine(190,60)(230,20)
\put(75,85){$g (p_g)$}
\put(75,35){$b (p_b)$}
\put(235,85){$W^- $}
\put(225,35){$t (p_t, s_t)$}
\put(160, 45){$b$}
\put(150,00){$(a)$}
\Gluon(310,100)(350,80){5}{4}
\ArrowLine(350,80)(390,100)
\Vertex(350,80){1}
\ArrowLine(350,50)(350,80)
\Vertex(350,50){6}
\Photon(350,50)(390,20){5}{5}
\ArrowLine(310,20)(350,50)
\put(285,85){$g (p_g)$}
\put(285,35){$b (p_b)$}
\put(390,35){$W^- $}
\put(390,85){$t (p_t, s_t)$}
\put(360,60){$t$}
\put(350,00){$(b)$}
\end{picture}
\caption{ Feynman diagrams contributing to  $gb \to  tW^-$ 
production. 
The effective $tbW$ vertex is shown as a filled circle.}\label{feyngraph1}
\end{center}
\end{figure}
Because of the occurrence of the weak $tbW$ vertex, it is well suited
for the study of a possible anomalous $tbW$ coupling occurring because of
new physics effects beyond the SM.

We concentrate here on observables in the above single-top $W$ 
associated 
production process which are odd under naive time 
reversal (T). By the term naive time reversal we mean an operation 
limited to
reversing the signs of momenta and spins of all particles. By contrast,
genuine time reversal also involves interchange of initial and
final states, and is therefore a transformation difficult to achieve in
 actual practice. 
At tree level, unitarity implies that the forward and backward
amplitudes have the same phase, and therefore the time reversal
violation, if present,
would be genuine. It would therefore occur
only in the presence of CP violation, on account of the CPT theorem.
On the other hand if the amplitude has an absorptive part arising from
rescattering or loop diagrams, characterized by a nonzero imaginary 
part for an anomalous effective coupling, 
the forward and backward amplitudes are
not the same, and the subsequent T violation would not be a genuine one.
In that case the CPT theorem does not necessarily imply CP violation.

We study in the process $pp \to tW^- X$
three observables which are T odd: (1) The polarization of the $
t$ perpendicular to the production plane (the transverse
polarization). (2) An asymmetry between the cross sections with a
positive and a negative value of the combination 
which corresponds to the angular-variable part of the three-momentum 
vector triple product
$\vec p_{P}\times \vec p_{t} \cdot \vec p_{\ell^-}$, where $P$, $t$ and
$\ell^+$ correspond respectively to an initial proton, the top quark 
and the charged lepton produced in top decay. (3) The mean
value of this combination of angular variables.
These are studied in the presence of possible anomalous $tbW$ couplings
in an effective Lagrangian. It is customary to consider an effective
Lagrangian arising in what is termed as an effective field theory, 
where the SM Lagrangian is
supplemented by higher dimensional terms which are suppressed by inverse
powers of a cut-off scale, a scale up to which the effective field
theory is valid. However, here we do not make any such specific
assumption about scales, but simply write an effective Lagrangian
permitted by Lorentz invariance.

The parton-level process which contributes to the process of interest is
$g b \to t W^-$. While defining the T-odd variables for the parton-level
process is straightforward, they average out to zero for a symmetric
 $pp$ initial state at the LHC. We therefore 
make a choice for the $z$ axis to be along the
direction of the combined momentum of $t$ and $W^-$ in the laboratory
(lab) frame. The $t$ momentum is then chosen to be in the $xz$ plane,
with the $x$ component of the $t$ momentum taken as positive.

Though higher order corrections to the cross section are available,
these are not known for the asymmetries and mean values we calculate. 
We restrict ourselves to the lowest order calculations. It is likely
that since we are dealing with ratios of quantities, the results for the
observables will be reasonably accurate.

Some previous work on T-odd observables in processes involving the top 
quark may be found in 
\cite{Tiwari:2019kly, Tiwari:2022cva, Tiwari:2022pug, Rindani:2024lvh}. 
CP violation in the $tW^-$ and $\overline t W^+$ production processes
was studied in \cite{pankajCP}.

\section{\boldmath $tbW$ couplings: CP and T
properties}\label{couplings}
The Lagrangian for effective $tbW$ couplings may be written as
\begin{equation}\label{lag}
\begin{array}{rcl}
{\cal L}_{Wtb}& =& \displaystyle \frac{g}{\sqrt{2}} \left[ 
W_\mu \overline t \gamma^\mu (V_{tb} f_{1L} P_L + f_{1R} P_R)b \right.
 \\
 &&\left. \displaystyle  -\frac{1}{2 m_W} W_{\mu\nu}\overline t \sigma^{\mu\nu}
(f_{2L} P_L + f_{2R} P_R)b  \right] + {\rm H.c.}
\end{array}
\end{equation}
Here, $g$ is the semi-weak coupling, $P_{L,R}$ are the left (right) chirality projection matrices, and
$W_{\mu\nu} = \partial_\mu W_\nu - \partial_\nu W_\mu$.
We later assume $V_{tb} = 1$ for simplicity. In the SM at tree
level, $f_{1L}=1$, and the remaining $f_i$ are vanishing.

In general, the anomalous couplings 
$f_{1R}$, $f_{2L}$ and $f_{2R}$ can be complex. The presence of an
imaginary part can signal either an absorptive part in the form factor
or
violation of CP or both. Whether the coupling is CP violating or not can
be checked 
by a comparison of the phases in the conjugate processes involving the
$t$ and the $\overline t$. The CP-violating phases would be opposite in
sign to each other in these conjugate processes. 
The absorptive phase, however, would be the
same in both these processes. A discussion of these issues may be found
in \cite{bern,Antipin:2008zx,pankajCP}. We recapitulate
 the relevant
points below.
Anomalous couplings, including the absorptive parts, have been
evaluated in certain theoretical scenarios in ref. \cite{bern}.

We can write the most general 
effective vertices up to mass dimension 5 for the four distinct 
processes of 
$t$ decay, $\overline t$ decay, $t$ production and $\overline t$ 
production as 
 \be\label{tdecay}
 V_{t\to bW^+} =- \frac{g}{\sqrt{2}} V_{tb} [ \gamma^\mu 
        (f_{1L}^* P_L + f_{1R}^* P_R ) 
        -i(\sigma^{\mu\nu}q_\nu/m_W)  (f_{2L}^* P_L + f_{2R}^*
        P_R ) ],
\ee
\be
 \label{tbardecay}
 V_{\bar t\to \bar bW^-} = -\frac{g}{\sqrt{2}} V_{tb}^*[\gamma^\mu 
        (\bar f_{1L}^* P_L + \bar f_{1R}^* P_R ) 
        -i(\sigma^{\mu\nu}q_\nu/m_W) (\bar f_{2L}^* P_L + \bar f_{2R}^*
        P_R )],
\ee
\be
 \label{tprod}
 V_{b^*\to tW^-} = - \frac{g}{\sqrt{2}} V_{tb}^* [ \gamma^\mu 
        (f_{1L} P_L + f_{1R} P_R ) 
        -i(\sigma^{\mu\nu}q_\nu/m_W) (f_{2L} P_R +
        f_{2R} P_L ) ],
\ee
\be
 \label{tbarprod}
 V_{\bar b^*\to \bar tW^+} = - \frac{g}{\sqrt{2}} V_{tb} [ \gamma^\mu 
        (\bar f_{1L} P_L + \bar f_{1R} P_R ) 
        -i(\sigma^{\mu\nu}q_\nu/m_W) (\bar f_{2L} P_R +
        \bar f_{2R} P_L )],
 \ee
where $f_{1L}$, $f_{1R}$, $f_{2L}$, $f_{2R}$,
$\bar f_{1L}$, $\bar f_{1R}$, $\bar f_{2L}$, and $\bar f_{2R}$ are form
factors, $P_L$, $P_R$ are left-chiral and right-chiral projection
matrices, and $q$
represents the $W^+$ or $W^-$  momentum, as applicable in each case.
In the SM, at tree level, $f_{1L}=\bar f_{1L} = 1$, and all other form
factors vanish. New physics effects would result in deviations of
$f_{1L}$ and  $\bar f_{1L}$ from unity, and nonzero values for other
form factors.

At tree level, when there are no absorptive parts in the
relevant amplitudes which give rise to the form factors, the following
relations are obeyed: 
 \be\label{noabspart}
 \begin{array}{rccl}
 f_{1L}^* = \bar f_{1L};& f_{1R}^* = \bar f_{1R}; & f_{2L}^* = \bar
 f_{2R}; & f_{2R}^* = \bar f_{2L}.
 \end{array}
 \ee
This can be seen to follow from the fact that in the absence of
absorptive parts, the effective
Lagrangian is Hermitian. Thus, the Hermiticity of the Lagrangian
of eq. (\ref{lag})
from which the amplitudes
(\ref{tdecay})-(\ref{tbarprod}) may be derived implies the relations 
(\ref{noabspart}).
On the other hand, if CP is conserved, the relations obeyed by the form
factors are
\be\label{CP}
\begin{array}{rccl}
f_{1L} = \bar f_{1L};& f_{1R} = \bar f_{1R}; & f_{2L} = \bar
f_{2R}; & f_{2R} = \bar f_{2L}.
\end{array}
\ee

The phases of the form factors can be expressed as sums and 
differences of two phases, one corresponding to a nonzero absorptive
part, and another corresponding to CP nonconservation:
 \be\label{phases}
 \begin{array}{rl}
 f_{1L,R} = \vert f_{1L,R} \vert \exp{(i\alpha_{1L,R} + i\delta_{1L,R})};&
 \bar f_{1L,R} = \vert f_{1L,R} \vert \exp{(i\alpha_{1L,R} -
 i\delta_{1L,R})};\\
 f_{2L,R} = \vert f_{2L,R} \vert \exp{(i\alpha_{2L,R} + i\delta_{2L,R})};&
 \bar f_{2R,L} = \vert f_{2L,R} \vert \exp{(i\alpha_{2L,R} - i\delta_{2L,R})},\\
 \end{array}
 \ee
where, for convenience, magnitudes of the couplings are assumed to be
equal in appropriate pairs.
In case there are no absorptive parts ($\alpha_i = 0$), we get the
relations (\ref{noabspart}) as a special case. In case of CP
conservation, we get the relations (\ref{CP}) as a special case. In what
follows, we will deal with a more general case when both CP violation
and absorptive parts may be present. 

\section{T-odd observables}
We investigate three representative T-odd observables. The first one is 
the polarization of the top quark transverse to the
production plane. 
The polarization of top quarks may be measured using the kinematic
distributions of the decay particles, without contamination
from the hadronization process, since the quark being heavy 
is expected to decay
before hadronization. 

The polarization can be measured experimentally from the decay
distribution of one of the decay products in the rest frame of the
top quark.
The angular distribution of a decay product $f$ for a top
 ensemble has the form
 \begin{equation}
 \frac{1}{\Gamma_f}\frac{\textrm{d}\Gamma_f}{\textrm{d} \cos \theta _f}
=\frac{1}{2}(1+\kappa _f P_{t} \cos \theta _f).
 \label{topdecaywidth}
 \end{equation}
 Here $\theta_f$ is the angle between the momentum of fermion $f$ 
and the top spin
vector in the top rest frame and $P_{t}$
is the degree of polarization of the top ensemble.
$\Gamma_f$ is the partial decay width and $\kappa_f$ is the spin
 power of $f$. For a charged lepton $f \equiv \ell^-$ 
the analyzing power is $\kappa_{\ell^-} = -1$ at tree level. 
We do not discuss the details of such a polarization 
measurement. However, it is appropriate to note that a measurement of
the top polarization using the  angular distribution of
$\ell^-$ will not be affected by the contribution of an anomalous 
$tbW$ coupling in the top decay process to linear order, as shown in
earlier work \cite{theorem}.

The other two T-odd observables involve using the momentum
of a top decay particle and constructing a triple three-vector product
of its momentum with the incoming beam momentum and the $t$
momentum. Since these three-vectors change sign under T, such a 
triple product
would be odd under T. For reasons discussed later, we will use only 
the factors
involving angular variables and not the whole triple product.

All these observables, being T odd, depend on the existence of an
absorptive amplitude, as discussed earlier. In an approach with 
 an effective Lagrangian
like the one in eq. (\ref{lag}), this implies that the observable will be
proportional to the imaginary part of an effective coupling. In our
case, it turns out that Im~$f_{2R}$ is the only one which contributes.
While we could also analyze other observables which can measure the real
parts of the anomalous couplings, we do not consider them here, and 
we set the 
real parts of the anomalous couplings to zero.
 
In order to calculate the T-odd observables,
we first obtain analytic expressions for the
production spin density matrix elements.  
For calculating helicity amplitudes and
density matrix elements, 
we have made use of the algebraic manipulation software FORM \cite{verm}
 as well as well the procedure outlined in \cite{VegaWudka}.
We also calculated the transverse polarization of the top quark
using FORM.
We then evaluate numerically cross sections, expressions and asymmetries for
various different cm energies at which the LHC has been operational in  the
past or will be in the future, viz., 7,  8, 13 and 14 GeV.
We use only leading-order expressions, in the absence of detailed NLO
expressions for polarized processes, though NLO corrections to the
cross sections are known. It is  expected that since the asymmetries 
and 
expectation
values we calculate are ratios of cross sections, they may be somewhat
insensitive to the NLO corrections.

Though analytical expressions for all the elements of the spin 
density matrix $\rho_{ij}$,
where $i,j = 1,2$ refer
respectively to the helicities $+1/2$ and $-1/2$ of the $t$, 
have been obtained, and used in our calculations, they are somewhat 
lengthy. We give
below only the expression for a combination of off-diagonal elements on which
the T-odd observables calculated here depend. The full expressions will
be given elsewhere.

The expression for the production cross section in terms of the elements
 of $\rho$ is
\begin{equation}\label{sigmaSM}
\sigma =
\int dx_1\,dx_2\, g(x_1)b(x_2) \int 
d\cos\theta_{t}^{\rm cm} \,
\frac{|\vec p_{\,t}^{\,\rm cm}|}{16 \pi \hat s^{\frac{3}{2}} }
(\rho_{11} + \rho_{22}) + (x_1 \leftrightarrow x_2),
\end{equation}
where the superscript ``cm'' refers to the $g b$ cm frame
and where we employ the five-flavour scheme. 
Here, $\sqrt{\hat s}$ is the $gb$ energy in the cm
frame of the pair, and $\theta_{t}^{\rm cm}$ is the angle made by
the $t$ momentum with the $z$ axis, defined by the direction of the
momentum of the $tW^-$ final-state, which is the direction of the parton
($g$ or $b$) carrying the larger momentum fraction. 
$(x_1 \leftrightarrow x_2)$ denotes the expression obtained by an
interchange of $x_1$ and $x_2$ in the previous expression, where again
$\cos\theta_t^{\rm cm}$ is the polar angle of the $t$, defined with
respect to the appropriate $z$ axis.
For a gluon carrying a momentum fraction $x_1$ of the proton and the $b$
 quark carrying a momentum fraction $x_2$, or vice versa, 
the $pp$ cm energy is $\hat s = 4x_1x_2 s$. 
We neglect $m_b$ everywhere.

We next proceed to evaluate the three T-odd observables.

\section{\boldmath Transverse polarization of $t$}
The transverse top polarization, that is, the polarization transverse to
the top production plane, arises for nonzero Im~$f_{2R}$. 
As mentioned earlier, for a symmetric collider like the LHC, the 
forward and backward
directions cannot be distinguished. However, it is still possible to
define an asymmetry by choosing the $z$ axis as the direction of the 
the $tW^-$ pair in the parton cm frame. Thus, in the laboratory (lab.)
frame, the $z$ axis is in the direction of the parton which receives a
larger Lorentz boost to take it from the parton cm frame to the lab
frame, because that, by momentum conservation, 
 is the direction of the momentum of the $tW^-$ state.  
The transverse polarization, thus defined, is the asymmetry given
symbolically by
\begin{equation}
P_t =
\frac{
\sigma_{x_b>x_g} (s_{t}^{\,y} >0)-
\sigma_{x_b>x_g} (s_{t}^{\,y} <0)+
\sigma_{x_g>x_b} (s_{t}^{\,y} >0)-
\sigma_{x_g>x_b} (s_{t}^{\,y} <0)
}{ 
\sigma_{x_b>x_g} (s_{t}^{\,y} >0)+
\sigma_{x_b>x_g} (s_{t}^{\,y} <0)+
\sigma_{x_g>x_b} (s_{t}^{\,y} >0)+
\sigma_{x_g>x_b} (s_{t}^{\,y} <0) 
},
\end{equation}
where $x_b$ and $x_g$ refer
respectively to the momentum fractions carried by the $b$ quark and the
gluon. $\vec s_t^y$ is the $y$ component of the spin of the top, the $y$
axis being along $\vec p_b \times \vec p_t$ for $x_b>x_g$, the events 
when the $b$ quark
receives the larger boost. The $y$ axis is along $\vec p_g \times \vec
p_t$ when $x_g > x_b$, the events when the $g$ parton receives the larger boost.

The numerator of this asymmetry, which may be called 
 the polarized cross section $\sigma_{\rm pol.}$,
is an integral over the phase space of the squared matrix element for
the production of a top with polarization perpendicular to the
production plane. 
It is the net polarization
along the $y$ direction. The spin density matrix elements for the
production  of the top were evaluated using FORM \cite{verm}. The
result for the imaginary part of the off-diagonal density matrix
element is given
by the equation 
\begin{equation}
-i(\rho_{12}-\rho_{21}) =  -{\rm Im}~f_{2R} \frac{g^2g_s^2}{3m_W\hat s}
\epsilon(p_g,p_b,p_t,s_t) ,
\end{equation}
where $\epsilon(p_g,p_b,p_t,s_t)$ is the Levi-Civita tensor $\epsilon$
contracted with three momenta and the spin four-vector of the top,
chosen along the $y$ axis as
described above.
The polarized top cross section is then given by
\begin{equation}\label{polcs}
\sigma_{\rm pol.} =
\int dx_1\,dx_2\, g(x_1)b(x_2) \int 
d\cos\theta_{t}^{\rm cm} \,
\frac{|\vec p_{\,t}^{\,\rm cm}|}{16 \pi \hat s^{\frac{3}{2}} }
[-i (\rho_{12} - \rho_{22})] + (x_1 \leftrightarrow x_2),
\end{equation}
where, as before, 
$(x_1 \leftrightarrow x_2)$ denotes the expression obtained by an
interchange of $x_1$ and $x_2$ in the previous expression.

The denominator, which is the unpolarized cross section
$\sigma_{\rm unpol.}$, is precisely the cross section given in 
eqn. (\ref{sigmaSM}).
The transverse polarization, as anticipated, is proportional to 
Im~$f_{2R}$.
We have evaluated it to linear order in the imaginary parts of the 
anomalous couplings, neglecting the real parts.

We also considered the transverse polarizations arising from Im~$f_{1R}$
and Im~$f_{2L}$.
It turns out that they vanish. 
Thus, the measurement of a
non-vanishing transverse polarization would signal specifically 
the presence of a nonzero coupling Im~$f_{2R}$.

We evaluate the relevant integrals numerically.
We use the values $m_t = 172.5$ GeV, $\sin^2\theta_W = 0.223$, where $\theta_W$ is
the electroweak mixing angle,  $m_W = 80.379$ GeV, and, to be used 
in the following
section, $\Gamma_W = 
2.085$ GeV for the $W$ boson width. We employ CTEQL1 parton 
distributions with $Q=m_t$ in the five-flavour scheme.

To estimate how well an asymmetry $A$ can measure Im~$f_{2R}$ 
for a  certain integrated luminosity $L$, we 
assume that the number of asymmetric
events $\sigma_{\rm pol.}L = A\sigma_{\rm unpol.}L$ is larger than the  
$\sqrt{N_{\rm SM}}$,  
the statistical fluctuation in the SM events in the absence of 
Im~$f_{2R}$, thus
allowing the discrimination of the polarization arising from the
anomalous coupling from the statistical background at the 1-$\sigma$
level. This is expressed by the inequality
\begin{equation}
A \sigma_{\rm SM} L > \sqrt{\sigma_{\rm SM} L}. 
\end{equation}
Thus, the limit on Im~$f_{2R}$ that can be placed at the 1-$\sigma$
level with luminosity $L$ is
\begin{equation}\label{Alimit}
{\rm Im}~f_{2R}^{\rm lim} = \frac{1}{\sqrt{\sigma_{\rm SM}L} A_1},
\end{equation}
where $A_1$ is the value of the asymmetry for ${\rm Im}~f_{2R}=1$.

We assume for uniformity an integrated luminosity of 100 fb$^{-1}$ for
all energies. 
The results for the transverse polarization asymmetry for unit value of
coupling Im~$f_{2R}$ and the limits on the coupling for the cm energies
considered here and with an integrated luminosity of 
$L = 100~{\rm fb}^{-1}$ 
are given in Table 1, together with the cross section in the SM. As
mentioned earlier, we are only using tree-level expressions. It is
likely that the asymmetries being ratios of cross sections, 
will be insensitive to the higher-order corrections. Since the NLO cross
sections are higher, our estimate of the limits would tend to be
conservative.

\begin{table}[hbt]
\begin{centering}
\begin{tabular}{cccc}
\hline
$\sqrt{s}$ (TeV) & $\sigma_{\rm SM}$ (pb)&$P_1$ & Im~$f_{2R}^{\rm lim}$\\ 
\hline
7 & 4.86 &   $5.74 \times 10^{-2}$&$2.50\times 10^{-2}$\\
8& 7.11 &  $5.47 \times 10^{-2}$& $2.17\times 10^{-2}$\\
13& 24.8 &  $4.44 \times 10^{-2}$&$1.43\times 10^{-2}$\\
14& 29.5 &  $4.27 \times 10^{-2}$&$1.36\times 10^{-2}$\\
\hline
\end{tabular}
\caption{The top transverse polarization $P_1$ for unit Im~$f_{2R}$
and the limits Im~$f_{2R}^{\rm lim}$ on Im~$f_{2R}$ for various
cm energies 
for an integrated
luminosity of 100 fb$^{-1}$.}
\end{centering}
\end{table}

\section{T-odd observables from angular variables}
We now consider the two other T-odd observables arising when the
$t$ decay is taken into account.
If we include the top decay $t \to b 
\ell^+ \nu_{\ell}$, it is possible to construct T-odd
observables without explicit recourse to top spin. For example, at the
parton level,  a T-odd variable which may be constructed 
is $\epsilon_{\mu\nu\alpha\beta}\ p_g^\mu\ p_b^\nu\
p_t^\alpha\ p_\ell^\beta$.
It is a Lorentz-invariant quantity, and could be evaluated in any
frame. For example, in the parton cm frame
it evaluates to
 $ \sqrt{\hat s} |\vec p_b|| \vec p_{t}|E_l \sin\theta_{t}
\sin\theta_\ell \sin\phi_\ell $, with all quantities calculated in the
parton cm frame. 
However, as stated earlier, the charged-lepton angular distribution is
independent of the anomalous couplings in the decay vertex 
to first order \cite{theorem}. 
Hence, 
to isolate the effect of anomalous coupling Im~$f_{2R}$ in the
production process and to avoid the influence of anomalous couplings 
in decay, 
we consider only a combination of angular variables,
even though it is not Lorentz
invariant. We now construct two T-odd observables from 
\begin{equation}
O =  \sin\theta_{t} \sin\theta_\ell \sin\phi_\ell , 
\end{equation}
all variables being measured in the lab frame.
These are 

\noindent (i) $A(O)$, the asymmetry in the variable $O$ of the cross section.
This is the ratio 
to the total number of events
of the difference in the number of events for
which the $A(O)$ is positive and for which it is negative: 
\begin{equation}\label{Oasym}
A(O) =
\frac{
\sigma (O>0)
- 
\sigma (O<0)
}{ 
\sigma (O>0)
+ 
\sigma (O<0)
}.
\end{equation}
The asymmetry $A(O)$ and the corresponding limits on Im~$f_{2R}$ for
various cm energies are given in Table 2 for an integrated luminosity of
100 fb$^{-1}$.

\noindent (ii)
The expectation value $\langle O \rangle$ of the variable $O$. 

The contributions to these observables, as mentioned earlier, come
only from the imaginary parts of the couplings. Thus, only the
off-diagonal elements of $\rho$, already shown earlier,  are relevant, 
the diagonal elements
being real. 

To evaluate these observables, the density matrix for the production of
the top has to be combined with the 
density matrix elements for the decay $t \to  
b \ell^+ \nu_\ell$, 
with an integration over all variables except the charged-lepton 
angular variables $\theta_\ell^0$ and $\phi_\ell^0$. These are now
written in the top
rest frame. The expression for the full cross section  is evaluated 
using the narrow-width approximation for the $W$. 

The density matrix elements $\Gamma_{ij}$, $(i,j = 1,2)$ are
(see for example \cite{Godbole:2002qu})
\begin{equation}
\begin{array}{ccl}
        \Gamma_{11}& =& K(1 + \cos\theta_\ell^0)\\
        \Gamma_{22}& =& K(1 - \cos\theta_\ell^0)\\
        \Gamma_{12}& =& K\sin\theta_{\ell}^0 e^{i \phi_\ell^0}\\
        \Gamma_{21}& =& K\sin\theta_{\ell}^0 e^{-i \phi_\ell^0}
\end{array}
\end{equation}
where 
\begin{equation}
        K = g^4(m_t^2 - 2p_t\cdot p_\ell).
\end{equation}

In principle, the anomalous $tbW$ couplings would contribute to the
decay density matrix. However, restricting to linear order, the effect
of the
anomalous couplings on the normalized angular distributions has been 
shown to be absent \cite{theorem}. We therefore do not include anomalous
couplings, assuming them to be small enough to ignore the quadratic
powers. 

The cross section for the process in the lab frame is given
by 
 \begin{equation}\label{cs6lab}
\begin{array}{rcl}
 \sigma& =& \displaystyle \int dx_1\, g(x_1) \int dx_2\, b(x_2)
\int d\cos\theta_{t}^{\rm lab} \int d\cos\theta_{\ell^-}^{\rm
lab}\int d\phi_{\ell^-}^{\rm lab}
\int dE_{\ell^-}^{\rm lab} 
 \frac{E_{\ell}^{\rm lab} p_{t}^{\rm
cm}}{\hat s^{3/2}} 
\\ & \times&\displaystyle 
 \frac{1}{512 (2\pi)^4 \Gamma_t\Gamma_Wm_tm_W}
\displaystyle
 (\rho_{11}\Gamma_{11}+\rho_{22}\Gamma_{22} + 
\rho_{12}\Gamma_{12} + \rho_{21}\Gamma_{21}) + 
(x_1 \leftrightarrow x_2).
	\end{array}
 \end{equation}
Here, $\Gamma_t$ ($\Gamma_W$) are the total widths of the $t$ (
$W$), $b(x)$ is the $b$ parton distribution 
function
in the proton, $E_\ell^{\rm lab}$ is the energy of the decay lepton in
the lab frame, and $p_{t}^{\rm cm}$ is the magnitude of the
$t$ three-momentum  in the parton cm frame. It is understood that as
before the
$z$ axis is chosen along the direction of the parton momentum which has
the larger momentum fraction of the proton, and the two possibilities
for choice of $z$ axis are summed over.

The asymmetry $A(O)$ is evaluated by using the cross section of eqn.
(\ref{cs6lab}) in eqn. (\ref{Oasym}).
As in the case of the transverse polarization asymmetry, the sensitivity
of the event asymmetry of $O$ may be measured by the 1-$\sigma$ limit on
the coupling given by the same eqn. ({\ref{Alimit}), 
where now for $A_1$ we use $A(O)_1$, the
value of $A(O)$ for ${\rm Im} f_{2R}=1$.
In Table 2 we give the values of 
$A(O)_1$ 
and the limits Im~$f_{2R}^{\rm lim}$ on Im~$f_{2R}$ for various
cm energies 
for an integrated
luminosity of 100 fb$^{-1}$.
\begin{table}[hbt]
\begin{centering}
\begin{tabular}{cccc}
\hline
$\sqrt{s}$ (TeV) & $\sigma_{\rm SM}$ (pb)&$A(O)_1$ & Im~$f_{2R}^{\rm lim}$\\ 
\hline
7 & 0.495 &  0.257  & $1.75 \times 10^{-2}$\\
8& 0.723 & 0.273  & $1.36 \times 10^{-2}$ \\
13& 2.53 & 0.326  & $6.10 \times 10^{-3}$ \\
14& 3.01 & 0.332  & $5.49 \times 10^{-3}$ \\
\hline
\end{tabular}
\caption{The asymmetry $A(O)_1$ in the observable $O$ 
for unit Im~$f_{2R}$
and the limits Im~$f_{2R}^{\rm lim}$ on Im~$f_{2R}$ for various
cm energies 
for an integrated
luminosity of 100 fb$^{-1}$.}
\end{centering}
\end{table}

The expectation value of $O$, for the case when the gluon carries a
momentum fraction $x_1$ of the proton and the $b$ quark carries a
momentum fraction $x_2$, can be calculated using the expression in eqn.
(\ref{cs6lab}).
\comment{
is then 
 \begin{equation}
\begin{array}{rcl}
 \langle O \rangle &\!\!\! =\!\!\!& \displaystyle \frac{1}{\sigma}\int dx_1 g(x_1)
\int dx_2 b(x_2)
\int d\cos\theta_{t}^{\rm lab} \int d\cos\theta_{\ell^-}^{\rm
lab}\int d\phi_{\ell^-}^{\rm lab}
\int dE_{\ell^-}^{\rm lab}
 \frac{E_{\ell}^{\rm lab} p_{t}^{\rm
cm}}{\hat s^{3/2}} 
\\ & \times &\displaystyle 
 \frac{1}{512 (2\pi)^4 \Gamma_t\Gamma_Wm_tm_W}
 (\rho_{11}\Gamma_{11}+\rho_{22}\Gamma_{22} + 
\rho_{12}\Gamma_{12} + \rho_{21}\Gamma_{21})\,O +
(x_1 \leftrightarrow x_2).
	\end{array}
 \end{equation}
} 
As in the case of the transverse polarization, it turns out the
observable $\langle O \rangle $  is zero in the case of the the anomalous
couplings $f_{1R}$ and $f_{2R}$.

For the observable $O$ to be measurable at the 1-$\sigma$ level with an
integrated luminosity $L$, its
expectation value should satisfy
\begin{equation}
\vert \langle O \rangle -  \langle O \rangle_{\rm SM} \vert 
> \frac{\sqrt{ \langle O^2 \rangle_{\rm SM} - 
\langle O \rangle_{\rm SM}^2}}{\sqrt{\sigma_{\rm SM} L}}.
\end{equation}
For our case, $\langle O \rangle_{\rm SM} = 0$. 
The 1-$\sigma$ limit on the coupling Im~$f_{2R}$ is then given by 
\begin{equation}
{\rm Im}~f_{2R}^{\rm lim} =
\frac{\sqrt{ \langle O^2 \rangle_{\rm SM}  
}}{\left(\langle O \rangle_{{\rm Im} f_{2R} = 1}
\right)\sqrt{\sigma_{\rm SM} L}}.
\end{equation}
The expectation values
and the limits on Im~$f_{2R}$ for various cm energies are given in 
Table 3, assuming an integrated
luminosity of 100 fb$^{-1}$.
\begin{table}[htb]
\begin{centering}
\begin{tabular}{ccccc}
\hline
$\sqrt{s}$ (TeV) & $\sigma_{\rm SM}$ (pb)&$\langle O \rangle_1$  
&$\langle O^2 \rangle_{\rm SM}$ & Im~$f_{2R}^{\rm lim}$\\ 
\hline
7 & 0.495 & $5.25\times 10^{-3}$ &$9.81\times 10^{-2}$ & $0.268$\\
8& 0.723 & $4.73\times 10^{-3}$ &$9.51\times 10^{-2}$ & $0.242$ \\
13& 2.53 & $3.06\times 10^{-3}$ &$8.51\times 10^{-2}$ & $0.190$ \\
14& 3.01 & $2.83\times 10^{-3}$ &$8.37\times 10^{-2}$ & $0.186$ \\
\hline
\end{tabular}
\caption{The SM cross section for $t$ production and decay
into the leptonic channel, the expectation value of $O$ for unit 
coupling, the expectation value of $O^2$ in
the SM, 
and the limits Im~$f_{2R}^{\rm lim}$ on  Im~$f_{2R}$ for various cm
energies, for an integrated
luminosity of 100 fb$^{-1}$.}
\end{centering}
\end{table}
The limits in Table 2 obtainable from the observable $O$ are comparable
to those in Table 1 which would result from the transverse polarization,
The limits from $\langle O \rangle$ are somewhat worse. However, it 
would not be fair to make
comparisons as the systematics are different for the different
measurements and a more realistic study of the measurements of various
momenta is needed. A more detailed study of the kinematics and
relevant cuts for particle identification would be required to get more
accurate sensitivities. However, our numbers are expected to 
 give a reasonable indication of the true values.
 
\section{Conclusions and Discussion}

We have considered three typical T-odd observables which might be 
measured
in the process $p p \to t W^- X$ in an extension of the
SM with anomalous $tbW$ interactions. 
Use has been made of analytical expressions for the spin density matrix
for the production and, when needed, the decay of the top quark to
calculate the cross section and the T-odd observables. 
These T-odd observables are
dependent on the imaginary part of the off-diagonal elements of the
density matrix, and hence, the imaginary part of the anomalous
couplings. 
Of the three anomalous couplings $f_{1R}$, $f_{2L}$ and $f_{2R}$, only 
the
imaginary part of $f_{2R}$ gives rise to  nonzero
values for these three observables.
Thus, a measurement of the T-odd observables we suggest here would
serve to isolate the coupling Im~$f_{2R}$ from amongst the three
anomalous couplings available.

The 1-$\sigma$ limits on Im~$f_{2R}$ we derive are of the order of 
10$^{-2}$, with for the T-odd asymmetries, and a little larger for the
expectation value of $O$. The limit improves for 
higher cm energy for the same integrated luminosity. 
Our calculation is restricted to a single channel for the 
top decay. Extending it  to the other two leptonic channels would 
improve the sensitivity.
 
It would be worthwhile to refine he sensitivities for the measurement 
of Im~$f_{2R}$ estimated here.
The measurement of the detailed final state is a difficult job for the
LHC detectors.
A more reliable estimate would require a detailed
study of the detector efficiencies and imposing of realistic kinematic
cuts for identification of the particles like the $t$, 
which is not attempted here.
It might be useful to resort to machine learning techniques which 
have been discussed for top tagging in recent times
\cite{Sahu:2023uwb, Kasieczka:2019dbj}.
Such a study would be worthwhile to carry out.

One may construct other more complicated T-odd observables, which are
products of the observable considered above with kinematic constructs
which are T-even, so that the product is odd under T. It may be possible
to optimize the sensitivity with the use of such a combination.

We do not consider hadronic top decays, which have a larger probability,
 since it would be difficult, if
not impossible, to measure a triple product of momenta, which would
require measurement of the charges of the quarks. However, given the
advantage of an enhanced sample of events in such a case, it would be
worthwhile to attempt using hadronic decays as well.

We have discussed naive T violation, not necessarily accompanied by CP
violation. We could also combine conjugate production and decay channels
to test and measure CP violation, if it exists. In principle, 
it would bring in more parameters, since the couplings $\overline
f_{iL}$ or $\overline f_{iR}$ associated with the $\overline t W^+$ 
production
process are independent of $f_{iL}$ and $f_{iR}$. However, if some
assumptions are made, as for example, those suggested in
Sec.\ref{couplings}, the $tW^-$ and $\overline tW^+$ production results
could be combined to test CP. In case CP conservation is assumed, the
assumptions would lead to increase in statistics for measurement of the
absorptive part.

It would also be interesting to study actual realizations of the T-odd
observables studied here in actual extensions of the SM, as for example
two-Higgs doublet models.

\noindent {\bf Acknowledgment} 
Support from the 
Indian National Science Academy, New
Delhi, under the Senior Scientist Programme 
is gratefully acknowledged.

	\end{document}